\newcommand{\be}{\begin{equation}}      
\newcommand{\ee}{\end{equation}}      

\documentclass[,final]            
  {aipproc}

\layoutstyle{6x9}

\begin{document}

\title{Real space statistical properties
of standard cosmological models}

\author{Andrea Gabrielli}{
  address={Dipartimento di Fisica, Universit\`a di Roma ``La Sapienza'',
P.le Aldo Moro 2, 00185 - Roma, Italy}
}

\author{Michael Joyce}{
  address={Laboratoire de Physique Nucl\'eaire et de Hautes Energies,
 Universit\'e de Paris VI, 4, Place Jussieu,
Tour 33 -Rez de chaus\'ee, 75252 PARIS Cedex 05}
}

\author{Francesco Sylos labini}{
  address={Laboratoire de Physique Th\'eorique,
         Universit\'e de Paris XI, B\^atiment 211,
  91403 Orsay, France}
}

\begin{abstract}
 After reviewing some basic relevant properties of stationary
 stochastic processes (SSP), we discuss the properties of the
 so-called Harrison-Zeldovich like spectra of mass density
 perturbations. These correlations are a fundamental feature of all
 current standard cosmological models. Examining them in real space we
 note they imply a {\it sub-poissonian} normalised variance in spheres
 $\sigma_M^2(R) \sim R^{-4} \ln R$. In particular this latter
 behaviour is at the limit of the most rapid decay ($\sim R^{-4}$) of
 this quantity possible for any stochastic distribution (continuous or
 discrete). In a simple classification of all SSP into three
 categories, we highlight with the name ``super-homogeneous'' the
 properties of the class to which models like this, with $P(0)=0$,
 belong. In statistical physics language they are well described as
 lattice or glass-like. We illustrate their properties through two
 simple examples: (i) the ``shuffled'' lattice and the
 One Component Plasma at thermal equilibrium.

\end{abstract}

\maketitle


\section{Introduction}

In standard theories of structure formation in cosmology the density
field in the early Universe is described as a perfectly uniform
and isotropic matter distribution, with superimposed tiny fluctuations
characterized by some particular correlation properties
(e.g. \cite{pee93}). These fluctuations are believed to be the initial
seeds from which, through a complex dynamical evolution, galaxies and
galaxy structures have emerged.  In particular, in all the standard
models, the initial fluctuations are taken to be Gaussian and with a
power spectrum (PS) $P(k)$ satisfying the so-called {\it Harrison-Zeldovich} (HZ)
condition of being proportional to $k$ at small $k$ \cite{har}.  
  
The present paper has two main porpuses \cite{prb}. Firstly, to
clarify the real space statistical properties of the mass density
fluctuations common to all the standard cosmological models
(which have been almost completely overlooked in the literature on the
subject).  And secondly, through this discussion, to relate and
compare these models of the primordial Universe to correlated systems
encountered in statistical physics.

In particular we find that all these standard cosmological models are
characterized by a ``superhomogeneous'' (or superuniform) matter
distribution. This means that mass fluctuations over sufficiently
large spatial scales are ``sub-poissonian'', i.e. they increase with
the spatial scale more slowly than in a random poissonian matter
distribution.  Moreover, it is shown that the scaling of mass
fluctuations satisfying the HZ condition approaches the slowest
possible for any stochastic matter distribution.  We will see that in
the context of usual statistical physics this kind of behavior is
recovered in the case of lattice-like or glass-like particle
distributions or in the so called One Component Plasma (OCP)
\cite{ocp} at thermal equilibrium. In this context we
present some simple recipe to build particle distributions satisfying
the HZ condition at small $k$.  This last point can be useful first of
all in the context of $N$-body simulations for the study of structure
formation from primordial fluctuations through a gravitational
dynamics \cite{thierry}. 

\section{Basic concepts of correlation analysis}

Inhomogeneities of the mass density field in cosmology are described
using the general framework of stationary stochastic processes
(hereafter SSP).  Let us consider in general the description of a
continuous or a discrete homogeneous mass distribution $\rho(\vec{r})$
in terms of such a process. A stochastic process is completely
characterized by its ``probability density functional'' ${\cal
P}[\rho(\vec{r})]$ which gives the probability that the result of the
stochastic process is the density field $\rho(\vec{r})$ (e.g. see
Gaussian functional distributions \cite{Vanmarcke}). For a discrete
mass distribution, the space (e.g. the infinite three dimensional
space) is divided into sufficiently small cells and the stochastic
process consists in occupying or not any cell with a point-particle
following certain correlated probabilities, and $\rho(\vec{r})$ can be
written in general as:
\[
 \rho(\vec{r})=\sum_{i=1}^{\infty} \delta(\vec{r} -\vec{r}_i)\,,          
 \]      
where $\vec{r}_i$ is the position vector of the particle $i$   
of the distribution .  

The word ``stationarity'' refers in the present context to the {\em
spatial} stationarity of the process, and means that the functional
${\cal P}[\rho(\vec{r})]$ is invariant under spatial translation.  We
suppose also that the distribution is {\em statistically isotropic}
(invariance of ${\cal P}[\rho(\vec{r})]$ under spatial rotation), and
has a well defined positive average value:
\[
\left< \rho(\vec{r}) \right> = \rho_0 > 0\,,   
\]  
where $\left< ... \right>$ is the ensemble average over all the possible   
realizations of the stochastic process, i.e. the average over the functional   
${\cal P}[\rho(\vec{r})]$.  
Moreover it is usually assumed that ${\cal P}[\rho(\vec{r})]$ is {\em ergodic}.
This means that spatial averages in single infinite realization
coincide with the ensemble averages.

The quantity $\left<\rho(\vec{r_1})\rho(\vec{r_2})...\rho(\vec{r_l})\right>$  
is called the        
{\em complete} $l$-point correlation function. In the discrete  
case $\left<\rho(\vec{r_1})\rho(\vec{r_2})...\rho(\vec{r_l})    
\right>dV_1,dV_2,...,dV_l$ gives the {\em a priori} probability of finding   
$l$ particles, in a single realization, placed in the infinitesimal
volumes $dV_1,dV_2,...,dV_l$ respectively around $\vec{r_1},
\vec{r_2},...,\vec{r_l}$.  In both cases the statistical stationarity
and isotropy (hereafter SSI) imply that the $l$-point correlation
functions, for any $l$, depend only on the scalar relative distances
among the $l$ points \cite{gsl01}.

Let us analyze in further detail the auto-correlation properties of these   
systems. As aforementioned, in our hypothesis,
$\left<\rho(\vec{r_1})\rho(\vec{r_2})\right>$ depends only on       
$r_{12}=|\vec{r_1}-\vec{r_2}|$. 

The {\em reduced} two-point correlation function 
$\tilde\xi(r)$ is defined by:
\[
\left<\rho(\vec{r_1})\rho(\vec{r_2})\right> \equiv   
\rho_0^2\left[1+\tilde\xi(r_{12})\right]\,.
\]
The correlation function $\tilde \xi(r)$ is one way to measure the
"persitence of memory" of spatial variations in the mass density
\cite{huang}.
For a discrete distribution of particles this means that $\rho_0\tilde
\xi(r)d^3 r$ measures the {\em excess} of probability of finding a particle
in a small volume $d^3 r$ at a distance $r$ from another fixed
particle, with respect to the {\em a priori} probability $\rho_0 d^3
r$. Obviously, since $\rho_0$ must be the average density in any
single realization in the infinite volume limit, $\tilde \xi (r)$
satisfies the condition of vanishing in the limit 
$r\rightarrow\infty$.

\subsection{The Poisson particle distribution}
Let us consider the paradigm of a stochastic homogeneous point-mass
distribution: the {\it Poisson case} (see \cite{gsl01}).  It can be
defined as follows: let us partition the space in cubic cells of
volume $\Delta V$, and then occupy each cell independently of the
others, with a point-particle with a probability $n_0\Delta V$, where
$n_0>0$ and $\Delta V\ll n_0^{-1}$. It is simple to show that in the
limit $\Delta V\rightarrow 0$, one can write:
\begin{eqnarray}
&&\rho_0=n_0\\
&&\tilde\xi(r)=\frac{\delta(\vec{r})}{\rho_0}\,
\label{poi1}
\end{eqnarray}
where $\delta(\vec{r})$ is the usual Dirac delta function.
Equation (\ref{poi1}) is a direct consequence of the fact that there is
no correlation between different spatial points in the definition of
the stochastic process.  That is, the reduced correlation function
$\tilde\xi$ has only the so called {\em diagonal} part.  It is simply
shown that this diagonal part is present in the reduced two-point
correlation functions of any statistically homogeneous discrete
distribution of particles with correlations.
Therefore in general \cite{Landau} for a  
SSI distribution of particles the reduced    
correlation function can be written as    
\[
\tilde\xi(r)=\frac{\delta(\vec{r})}{\rho_0}+\xi(r)
\]
where $\xi$ is the non-diagonal parts which is meaningful only for
$r>0$.

\section{The mass variance in a sphere}

In this section we consider the amplitude of the mass fluctuations   
in a generic sphere of radius $R$ with respect to the average mass.   
First let  $M(R)=\int_{C(R)}\rho(\vec{r}) d^3r$  
be the mass (for a discrete distribution  
the number of particles) inside the sphere $C(R)$ of radius $R$  
and then volume $\|C(R)\|=\frac{4\pi}{3}R^3$.   
The normalised mass variance is defined as     
\[
\sigma_M^2(R)=      
\frac{\langle M(R)^2 \rangle - \langle M(R) \rangle^2}   
{\langle M(R) \rangle^2}\,.     
\]
It is simple to show that $\sigma_M^2(R)$ in a stationary stochastic mass 
density can be rewritten as:
\be    
\label{v6}    
\sigma_M^2(R) =  
\frac{1}{\|C(R)\|^2} \int_{C(R)} d^3r_1\int_{C(R)} d^3r_2   
\tilde\xi(|\vec{r_1} - \vec{r_2}|)  \;.  
\ee   
This formula will be useful in the following for a
complete classification of the mass density fields with respect to the
large scale behavior of mass fluctuations.

Note that since $\tilde \xi(r)\rightarrow 0$ for $r\rightarrow \infty$, then
\[
\lim_{R \rightarrow \infty}  
\sigma_M^2(R) = 0 \;. 
\]  
This is nothing but the condition of existence of a well defined 
average density $\rho_0$.

\subsection{A classification of mass fluctuations}

If we apply Eq.~(\ref{v6}) to the case of a Poisson point process as
above defined, we find 
\[
\sigma_M^2(R) = \frac{1}{\rho_0\|C(R)\|}\equiv \frac{1}{\left<M(R)\right>}  \;,  
\] 
that is $\left<\Delta M^2(R)\right>\simeq\|C(R)\|$.  The same
widespread behavior is found in systems characterized by mainly
positive short range correlations (e.g. a perfect gas at high
temperature).

\begin{figure}
  \includegraphics[height=.4\textheight,angle=90]{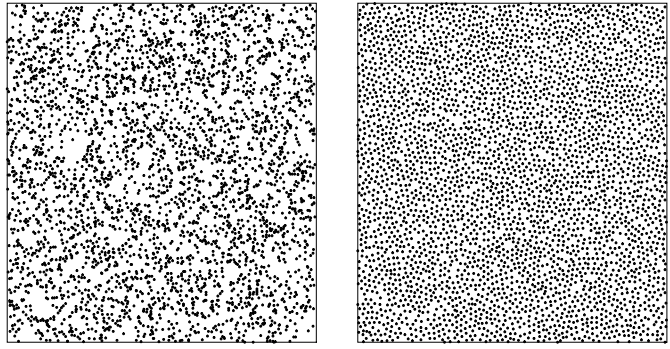}
  \caption{On the left: Poisson particle distribution ($10^4$ particles).
On the right: {\em superhomogeneous} particle distribution ($10^4$ particles).
Note the higher order level in the latter than in the former.}
\label{fig1}
\end{figure}

We can characterize an arbitrary mass distribution with
respect to the behavior of mass fluctuations as follows. Let us suppose
$\sigma_M^2(R)\sim R^{-\alpha}$ at large $R$, then:
\begin{itemize}
\item If $0<\alpha<3$ the system has {\em critical} fluctuations, typical
of the order parameter of a thermodynamical system at the critical
point of a second order phase transition. As we show below, this
happens in the case of mainly positive {\em long range} correlations
such that $\int d^3r \tilde\xi(r)=+\infty$;
\item If $\alpha=3$ we have a {\em substantially} poissonian system, 
characterized by mainly positive {\em short} range correlations, such that  $\int
d^3r \tilde\xi(r)=c$ where $0<c<+\infty$;
\item If $\alpha>3$ the system has sub-poissonian fluctuations typical of
lattice or glass-like systems with an almost ordered arrangement of
mass perturbations (of particles in the discrete case), characterized
by a balance between positive and negative correlations, such that
$\int d^3r \tilde\xi(r)=0$. For this reason we call these systems 
{\em superhomogeneous}. We show also below that 
$\alpha<4$ ($\alpha<d+1$ in $d$ dimensions) for any stochastic mass distribution.
 As shown below all the standard cosmological models
for primordial mass fluctuations belong to this class with $\alpha=4$
(for a difference between a Poisson and a superhomogeneous distribution
see Fig.~\ref{fig1}).
\end{itemize}

\section{The power spectrum $P(k)$}
Much more used in cosmology than $\tilde\xi(r)$ is the equivalent
$k$-space quantity, the power spectrum $P(\vec{k})$ which is defined
as
\[
P(\vec{k})=\lim_{V\rightarrow\infty}
\frac{\left<|\delta_\rho(\vec{k})|^2\right>}{V}    
\]     
where $\delta_\rho(\vec{k})=\int_V d^3r e^{-i\vec{k}\cdot \vec{r}}
(\rho(\vec{r})-\rho_0)/\rho_0$ is the Fourier integral in the volume
$V$ of the normalized fluctuation field
$(\rho(\vec{r})-\rho_0)/\rho_0$.  In a SSI matter distribution this
depends only on $k=|\vec{k}|$ and can be written as
\be   
P(\vec{k}) \equiv P(k) = \int d^3r e^{-i\vec{k}\cdot \vec{r}}
\tilde\xi({r})=4\pi\int_0^{\infty} dr\, r^2
\frac{\sin (kr)}{kr} \tilde\xi({r})\,.    
\label{lat7a}      
\ee
It follows from its definition that $P(k)\ge 0$ (this property and
Eq.~(\ref{lat7a}) constitute the so-called Kh$\hat{\mbox{\i}}$ncin
theorem of SSP's).  In particular for a Poisson particle distribution
one has $P(k)=\rho_0^{-1}$.  At last note the important relation
$P(0)=\int d^3 r \tilde \xi(r)$ for any SSP.

\subsection{The cosmological power spectrum, and the HZ condition}

In current cosmological models it is generally assumed that the power
spectrum $P(k) \sim k$ at small $k$. This is the famous
Harrison-Zeldovich (HZ) condition.  It is believed to describe the
``primordial'' fluctuations at very early times, the putative remnants
of a period of ``inflation'' prior to the ordinary Big Bang phase
\cite{pee93,padm}.  The HZ behaviour is appropriately cut-off at a short
distance scale (i.e. for $k$ larger than some cut-off $k_c$). The
type of the cut-off depends on the single cosmological model (e.g. Cold
Dark Matter, Hot Dark Matter, etc.).  The reason for the HZ condition
is tied to considerations about the consistency of the
Friedmann-Robertson-Walker (FRW) metrics under perturbations (see
\cite{pee93,padm}): any other spectrum will give mass fluctuations
which dominate over the homogeneous background either at some time in
the future or past.

One can see \cite{prb} that in all the standard cosmological models the
corresponding $\tilde\xi(r)$ is positive at small $r$ and has a negative 
tail $\sim -r^{-4}$ at large $r$ with only one zero. 

\subsection{The power spectrum vs. the mass variance}

Quite generally, by studying Eq.~(\ref{v6}), it is not difficult to
show
\cite{prb} that for a spectrum $P(k) \sim k^n$ for $k \rightarrow 0$
(note that the Poisson-like system are given by $n=0$) and
appropriately cut-off at large $k$, one has at large $R$
\be 
\label{pk-sr}  
\sigma_M^2(R) \sim   
\left\{ \begin{array}{lll} 
 1/R^{3+n} \; \;  \mbox{if} \;\; n<1\\   
\log(R)/R^{4} \;\; \mbox{if} \;\;  {n=1}  \\  
1/R^{4}  \;\; \mbox{if} \;\;  {n>1} 
\end{array}  
\right. 
\ee  
In terms of the non-normalized quantity $\left<\Delta M^2(R)\right>
\sim \sigma_M^2(R)R^6 $, Eq.~(\ref{pk-sr}) says that for $n>0$
(i.e. $\int d^3r \tilde\xi(r)=0$) we have a scaling behavior of
$\left<\Delta M^2\right>$ as a function of $R$ slower than for Poisson
fluctuations, which correspond to $n=0$ (i.e. $\int
d^3r\tilde\xi(r)=c>0$), with the limiting behavior for $n\ge 1$
corresponding to quadratic mass fluctuations which are proportional to
the {\it surface area} of the sphere. These systems are thus
characterised by surface fluctuations, ordered (or homogeneous enough,
one could say) to give this very particular behaviour.  The case $n<0$
(i.e. $\int d^3r
\tilde\xi(r)=+\infty$) is typical of critical phenomena in which long
range mainly positive correlations determine a ``super-poissonian''
behavior of quadratic fluctuations.

Equation (\ref{pk-sr}) shows clearly that cosmological models,
characterized by the HZ condition ($n=1$), are not only
superhomogeneous, but are at the limit of the slowest possible behavior
of quadratic mass fluctuations with the spatial scale.

\section{Two ``superhomogeneous'' examples} 

In this section we present two examples of superhomogeneous point-particle 
distribution: the shuffled lattice \cite{prb}
and the One Component Plasma (OCP) \cite{ocp}.

{\bf 1)} In the case of particles placed  
on the sites of a regular cubic lattice (in any dimension $d$), one
has \cite{prb} $\rho_0=a^{-d}$ where $a$ is the lattice spacing and:
\[
P_l(\vec{k})=(2\pi)^d\sum_{\vec{H}\ne 0}\delta(\vec{k}-\vec{H})\,,
\]
where the sum is extended to all the vector of the reciprocal lattice
but the origin $\vec{H}=\vec{0}$. Therefore we can say that around
$k=0$ $P_l(k)\sim k^{+\infty}$, and then it is the {\em most superhomogenehous}
particle distribution with $\sigma_M^2(R)\sim R^{-(d+1)}$.

At this point let us introduce a stochastic displacement field, in
which each particle is displaced from its lattice site $\vec{R}$ of a
random and statistically isotropic vector $\vec{u}_{\vec{R}}$, each
particle independently of the others.  One can show \cite{preprint}
that, if the probability density of the displacements $p(u)$ has a
finite variance, then the resulting particle distribution has a power
spectrum $P_{sl}(k)\sim k^2$ at small $k$.  If instead, the variance
is infinite, and $p(u)$ has a tail going as $u^{-a}$ (with $d<a<d+2$
for normalizability, but with infinite variance), then
$P_{sl}(k)\sim k^{a-d}$ at small $k$. Therefore, choosing
appropriately $a$ we can obtain any kind of superhomogeneous particle
distribution with the exponent of $P_{sl}(k)$ at small $k$ ranging
from $0$ to $2$.  In particular for $a=d+1$ ($a=4$ in three dimension)
we obtain the HZ condition.

{\bf 2)} Let us now present briefly another model taken from
equilibrium statistical physics displaying superhomogeneous mass
fluctuations: the OCP. It is a system of interacting point particles
carrying a unit positive charge and mass repulsing each other via a
Coulomb potential $V(r)=1/r$, in a continuous uniform background
negatively charged giving overall charge neutrality \cite{ocp}. One
can show quite generally that, at thermodynamic equilibrium the system
has only one phase, and at small $k$ the power spectrum of the
particle distribution satisfies:
\[
P(k) \sim \frac{k^2}{4\pi \rho_0^2 \beta}\,,
\]
where $\rho_0$ is the average density of particles and $\beta$ is the
usual inverse temperature. 
It is very interesting to note that \cite{lebo}
if, instead of having the Coulomb 
potential $V(r)\sim r^{-1}$, one has a potential $V(r)\sim r^{-2}$ one 
would obtain 
\[
P(k)\sim \frac{k}{2\pi^2 n^2 \beta}\,,
\]
that is the HZ condition. These results can be very important if applied
to $N$-body simulations of gravitational dynamics to study structure
formation starting from the primordial mass fluctuations
\cite{thierry}. In fact, in this context a great importance is given to
the accurate preparation of the initial conditions for the
simulation. For instance, this can be set through a modified OCP
with $V(r)\sim r^{-2}$ at large scales and an appropriate modified
behavior at smaller scales \cite{lebo}.

Both {\bf 1)} and {\bf 2)} provide particle distributions similar to
the right side picture of Fig.~\ref{fig1}.

\end{document}